\documentclass[prd,aps,superscriptaddress,10pt,nobibnotes,notitlepage,twocolumn,longbibliography,nofootinbib,amsmath,amssymb]{revtex4-1}
\usepackage{url}
\usepackage{hyperref}
\hypersetup{
    colorlinks=true,
    linkcolor=blue,
    filecolor=magenta,      
    urlcolor=blue,
    pdftitle={field-theory-axiverse},
    pdfpagemode=FullScreen,
    }

\usepackage{hhline}
\usepackage{enumitem}
\usepackage{subfigure}
\usepackage{graphicx}
\usepackage{dcolumn}
\usepackage{nicematrix}
\usepackage{slashed}
\usepackage{bm}
\def\beq{\begin{equation}}
\def\eeq{\end{equation}}
\def\be{\begin{equation}}
\def\ee{\end{equation}}
\def\bea{\begin{eqnarray}}
\def\eea{\end{eqnarray}}

\def\tl{\tilde{\lambda}}

\begin{document}
\title{The Field Theory Axiverse} 
\medskip\
\author{Stephon Alexander}
\email[Email: ]{ stephon\_alexander$@$brown.edu}
\affiliation{Department of Physics,
Brown University, Providence, RI 02912, USA}
\author{Tucker Manton} 
\email[Email: ]{ tucker\_manton$@$brown.edu}
\affiliation{Department of Physics,
Brown University, Providence, RI 02912, USA}
\author{Evan McDonough} 
\email[Email: ]{ e.mcdonough$@$uwinnipeg.ca}
\affiliation{Department of Physics, University of Winnipeg, Winnipeg MB, R3B 2E9, Canada}

\begin{abstract}
Axion and axion-like particles (ALPs) are a prominent candidate for physics beyond the Standard Model, and can play an important role in cosmology, serving as dark matter or dark energy, or both, drawing motivation in part from the string theory axiverse. Axion-like particles (ALPs) can also arise as composite degrees of freedom following chiral symmetry breaking in a dark confining gauge theory, analogous to the Standard Model (SM) pion. A dark sector with arbitrary $N_f$ flavors of dark quarks leads to $N_f^2-1$ axion-like states, effectively a {\it field theory axiverse} (or `$\pi$-axiverse'). A portal to the visible sector can be achieved through the standard kinetic mixing between the dark photon and SM photon, generating millicharges for the dark quarks and consequently couplings, both parity-even and parity-odd, between the SM and pions. This scenario has been studied for the $N_f=2$ case and more recently for $N_f=6$. In this work, we study the spectrum of this field theory axiverse for an arbitrary number of flavors, and apply this to the example $N_f=10$. We calculate the couplings to the SM photon analogous to the conventional axion-photon coupling, including the $N_f$ and $N_c$ dependence, and compute the present and future constraints on the $N_f=10$ $N_c=3$ $\pi$-axiverse. We elucidate the accompanying `bary-verse' of superheavy dark baryons, namely an ensemble of charged and neutral dark baryons with a mass set by the dark pion decay constant.
\end{abstract}

\maketitle

\tableofcontents

\section{ Introduction}

The microscopic nature of dark matter is one of the most significant puzzles in physics to date. Amongst many possibilities (see e.g.~\cite{Safdi:2022xkm} for a review) a prominent dark matter candidate is the {\it axion}, or the more general axion-like particle (ALP). Initially motivated by the strong CP problem of the Standard Model (SM) \cite{Peccei:1977hh,Wilczek:1977pj,Weinberg:1977ma}, the axion was shortly thereafter proposed as a cold dark matter candidate  \cite{Preskill:1982cy,Abbott:1982af,Dine:1982ah}. Axions and ALPs play a starring role in modern cosmology, where they can serve as not only the observed dark matter, but also provide a candidate for the cosmic inflation field  or the dark energy field (see \cite{Marsh:2015xka} for a review of axion cosmology).

String theory predicts tens to hundreds of ALPs \cite{Svrcek:2006yi,Arvanitaki:2009fg,Cicoli:2012sz}, whose properties are sensitive to the geometry and topology of the extra dimensions and the extended objects (eg. D-branes) and fields contained therein. The term `axiverse' \cite{Arvanitaki:2009fg} has been used to describe the spectrum of string theory axions, whose masses are expected to span as many as 30 orders of magnitude. This work will focus on an alternative path to an axiverse.

Since ordinary matter largely originates from quantum chromodynamics (QCD), it is natural to consider a dark version of the strong force such as dark QCD (dQCD) (see e.g.~\cite{Bai:2013xga,Tsai:2020vpi,Alexander:2020wpm}). This is appealing for many reasons but perhaps most importantly, QCD is well understood both theoretically and experimentally. Furthermore, dQCD with ultralight (dark) quarks contains a spectrum of ALPs with properties similar yet discernible from standard axion models. These states arise as pseudo-Nambu-Goldstone bosons (PNGBs) from the breaking of a global $SU(N_f)\times SU(N_f)$ symmetry, where $N_f$ is the number of dark quark flavors. The ALPs are composite degrees of freedom analogous to the SM pion, and have thus been dubbed \textit{$\pi$-axions} \cite{Alexander:2020wpm}. The
chiral symmetry is broken below the dQCD scale, $\Lambda_{\text{dQCD}},$ which arises as a result of dimensional transmutation in the same fashion as QCD in the SM \cite{Coleman:1973jx}; $SU(N_f)_{\text{L}}\times SU(N_f)_{\text{R}}\rightarrow SU(N_f)_{\text{V}}$ results in $N_f^2-1$ PNGBs, where $N_f-1$ states are real pseudoscalars which experience an axion-like coupling to the SM photon. For large $N_f$, this is effectively an axiverse completely independent of the string axiverse, which we will call the \textit{field theory axiverse}.

A dark matter construction of this type was first discussed and proposed in \cite{Maleknejad:2022gyf}; a further analysis was carried out in \cite{Alexander:2023wgk} exploring the cosmology and detection prospects in the specific case of a dark Standard Model, with $N_f=6$ as in the visible Standard Model. Compared to the many dark matter models on the market, including `dynamical' axion models such as \cite{Kim:1984pt,Choi:1985cb,Hook:2014cda,Gherghetta:2016fhp,Croon:2019iuh,Gaillard:2018xgk}, composite ultralight dark matter has been studied relatively sparsely \cite{Alexander:2018fjp,Alexander:2020wpm,Maleknejad:2022gyf,Alexander:2023wgk,Abe:2024mwa}. However, due to the significant efforts going into direct and indirect detection of axions and ALPs (see \textit{e.g.} \cite{Marsh:2015xka} for a review), it is particularly appealing to consider models that predict axion-like couplings to the SM photon which can be distinguished from the string axiverse and the standard QCD axion.

The field theory axiverse is characterized by a tightly packed mass spectrum for all $N_f-1$ real pseudoscalars, in contrast with the {\it logarithmic} distribution of axion masses in the string theory axiverse \cite{Broeckel:2021dpz}. Moreover, this axiverse has a {\it single decay constant}, again in contrast with the logarithmic distribution of decay constants in the string theory axiverse \cite{Broeckel:2021dpz}. This crucial difference owes to the fact that the field theory axiverse relies a single non-perturbative effect, namely confinement in the dark QCD, whereas the string theory axiverse utilizes an array of non-perturbative effects, roughly one-per-axion.

The field theory axiverse also exhibits couplings to the Standard Model, in particular the familiar ALP interaction with photons. The pion-photon interaction has strength $g_{\pi \gamma \gamma} \simeq {\cal A} N_c\varepsilon^2 \alpha_{\rm em}/F_{\pi}$, where ${\cal A}={\cal O}(1)$ is an anomaly coefficient, $\varepsilon$ is the millicharge parameter, $F_{\pi}$ is the decay constant. Comparison to the axion-photon coupling of single-field ALP dark matter, this can be understood as an  overall enhancement of the latter, sensitive to both the number of flavors $N_f$ and the number of colors $N_c$ of the dark QCD theory.

The structure of this letter is as follows. In Sec. \ref{sec:dSM}, we begin by presenting the model and discussing the relevant parameters and regimes of interest corresponding to ongoing axion detection efforts. We then dive into our primary results: in Sec. \ref{sec:Enumerating}, we derive the formulae which enumerate the axiverse states (charged, complex neutral, or real neutral) for an arbitrary number of flavors $N_f$, while in Sec. \ref{sec:MassSpectrum}, we review the Gell-Mann-Oakes-Renner relation, which approximates the mass of the $\pi$-axions. Sec. \ref{sec:AxionPhotonCoupling} is devoted to an analysis of an effective axion-photon coupling, which experiences an enhancement as a function of $N_f$ and $N_c$ that follows from the chiral anomaly and the presence of multiple real pseudoscalars. We then comment on the additional portals to the SM in Sec. \ref{sec:OtherPortals} before examining the specific case of $N_f=10,$ $N_c=3$ in Sec. \ref{sec:Nf10}, including the spectrum of the $10^2-1=99$ states in Sec. \ref{sec:Nf10Spectrum}, and experimental constraints in Sec. \ref{sec:Nf10Constraints}. Finally, in Sec. \ref{sec:Baryons}, we briefly discuss dark baryonic-like states composed of $N_c$ dark valence quarks. Such states are plentiful in models with arbitrary $N_f,$ $N_c$, and we present an approximate formula for enumerating this `bary-verse' associated to the field theory axiverse. We then conclude in Sec. \ref{sec:Discussion}, followed by a brief appendix where we write out the algorithm for computing the generators $T^a$ for an arbitrary $SU(N)$ (Appendix \ref{app:SUNgennies}), and the result of constructing the $\Sigma=T^a\pi_a$ matrix for $SU(10)$ (Appendix \ref{app:SigmaMatrix}).

\section{A Field Theory Axiverse}
\label{sec:dSM}

The focus of this work is a confining gauge theory, with $N_f$ flavors of dark quark and $SU(N_c)$ gauge group. The Lagrangian is given by  
\begin{equation}
\label{eq:dQCD}
    \mathcal{L}_{\mathrm{dQCD}}=-\frac{1}{2}{\rm Tr} \, G_{\mu\nu}G^{\mu\nu}+\sum_{i,j=1}^{N_f}\bar{q}^{i}(i\slashed{D}\delta_{ij}-m_{ij})q^{j},
\end{equation}
where $G_{\mu \nu}$ is the non-Abelian field strength tensor and $q^i$, $i=1...N_{f}$, denote Dirac fermions in the fundamental representation of the the dark $SU(N_c)$ gauge symmetry. The mass matrix $m_{ij}$ is in general an $N_f\times N_f$ matrix with off diagonal elements, however for simplicity we focus on the case of a diagonal mass matrix, and consequently do not allow for flavor changing currents. This is in difference from the dark Standard Model of Ref.~\cite{Alexander:2023wgk}, where ultra-heavy (dark) weak bosons were integrated out of the low energy spectrum to give effective vertices allowing for flavor changing processes. In this work, we assume only a dark strong sector, wherein all field content is neutral under Weak isospin.
Instead, the dark quarks are solely endowed with a millicharge under the SM $U(1)_{\text{em}}$, which arises through kinetic mixing between the dark and SM photon \cite{Cline:2021itd}. Moreover, that the dark quarks lack Weak isospin implies all anomaly cancellations are trivially satisfied (see \textit{e.g.} \cite{Schwartz:2014sze}).

This theory exhibits a global $U(N_f)\times U(N_f)$ chiral symmetry, which is broken at low energies, below the confinement scale $\Lambda_{\rm dQCD}$. In particular the breaking of the subgroup,
\begin{equation}
    SU(N_f)_{\text{L}}\times SU(N_f)_{\text{R}}\rightarrow SU(N_f)_V,
\end{equation}
leads to a spectrum of Goldstone bosons known as pions. In this work we consider the spectrum of pions as an effective `axiverse', analogous to the string theory axiverse \cite{Svrcek:2006yi,Arvanitaki:2009fg,Cicoli:2012sz}.

In its simplest form, this formulation of the axiverse has only two free dimensional parameters: the quark mass $m_q$ and the dark confinement scale $\Lambda_{\rm dQCD}$. At low energies, these determine the neutral  pion mass scale and pion decay constant as
\begin{equation}
    m_{\pi}^2 = m_q \Lambda_{\rm dQCD} \,\, \, , \,\,\, F_{\pi}\sim \Lambda_{\rm dQCD} .
\end{equation}
The present work focuses on the regime of a hierarchy,
\begin{equation}
    \frac{m_q}{\Lambda_{\rm dQCD}} \ll 1 ,
\end{equation}
in order to realize a pion mass scale and decay constant comparable to that conventionally associated with axions (see e.g. \cite{Marsh:2015xka}):
\begin{equation}
    m_{\pi} < {\rm  eV} \,\, , \,\, F_{\pi} \gtrsim 10^{11}\, {\rm GeV} .
\end{equation}
The dark pions constitute light axion-like particles, which can serve as dark energy, dark matter, or both. 

As a dark matter model, the two free parameters of the theory can be further reduced to 1 single free parameter
under the assumption that the dark pions constitute the entirety of the dark matter: Matching to the observed abundance of dark matter $\Omega_{\rm dm}h^2=0.12$ determines the decay constant $F_{\pi}$ as a function of the mass, leaving only a single free parameter, the mass scale $m_{\pi}$, for this dark matter scenario.

Remarkably, a portal to Standard Model may be achieved without introducing a new free parameter: we can endow our dark quarks with a charge under Standard Model identical to that of the SM quarks, namely, 
\begin{equation}
    Q_{e} = -\frac{1}{3}e,\frac{2}{3}e.
\end{equation}
In contrast with naive expectation, e.g. from experience with millicharged dark matter \cite{Cline:2021itd,Davidson:2000hf,Bogorad:2021uew,Alexander:2020wpm, Plestid:2020kdm}, the model is naturally safe from constraints on electrically charged dark matter due to the high confinement scale: the charge of the dark quarks is confined inside neutral pions up to a scale $\Lambda_{\rm dQCD} > 10^{11}$ GeV, and all charged states (e.g., charged dark pions, charged dark baryons) have mass at the dark QCD scale.

In this work, we will follow the millicharged dark matter convention and endow our dark quarks with a fractional electric charge parameterized by $\varepsilon \leq 1$. We moreover make a slight generalization of the one-third/two-third fractional charge, and only demand that the up- and down-type fractional charges satisfy
\begin{equation}
    Q_u-Q_d=\varepsilon.
\end{equation}
In what follows, we develop in detail the spectrum of pions, including their mass and coupling to the Standard Model.

\subsection{Enumerating the Axiverse}\label{sec:Enumerating}

For an arbitrary $SU(N_f)$ flavor symmetry breaking, one generally has $N_f^2-1$ states where $N_f-1$ are real scalars and electrically neutral. The remaining states are complex scalars, either electrically charged or neutral (analogous to the SM $\pi_\pm$ and neutral kaon $K^0$). In order to count the number of each states, one can imagine constructing a table where the columns are labeled by each quark charged, increasing by generation. Label each row by the charge of the anti-quark of the same flavor. In this table, the diagonal entries are related to the real, neutral states analogous to the SM $\pi_0$ and $\eta$ particle. Their specific quark substructure is dictated by the diagonal $SU(N_f)$ generators (\ref{diagGens}). The off diagonal states are all complex scalars whose composition can be identified using the generators (\ref{realoffGens}) and (\ref{imoffGens}), and their charge is given by the sum of the quark+anti-quark charge from the row/column in which it lies. An example of such a table is given in Table \ref{chargetable}.

\begin{table}[h]
\begin{tabular}{c|ccccc}
&$u_1^{Q_u}$ &$d_1^{Q_d}$ &$u_2^{Q_u}$ &$d_2^{Q_d}$ &$\cdots$  \\ \hline
$\bar{u}_1^{-Q_u}$ & 0 & -1 & 0  &-1 & $\cdots$  \\
$\bar{d}_1^{-Q_d}$ & 1 & 0 & 1 & 0 & $\cdots$ \\
$\bar{u}_2^{-Q_u}$ & 0 & -1  & 0 &-1&$\cdots$ \\
$\bar{d}_2^{-Q_d}$ & 1 &0 &1 &0 &$\cdots$ \\
$\vdots$ &$\vdots$ &$\vdots$ &$\vdots$ &$\vdots$& $\ddots$
\end{tabular}
\caption{\label{chargetable} We label each quark in increasing generation along the top row simply as $u_1,d_1,...$ along with their anti-quark pairs along the leftmost column. Their respective charges $\pm Q_u$ and $\pm Q_d$ are written in the superscripts.}
\end{table}
Since the states reflected over the diagonal are complex conjugates, we can focus on the states above the diagonal only. We see that the set of states that are the first diagonal above the primary diagonal, which start with the $d_1\bar{u}_1$ bilinear with charge -$\varepsilon$, are all charged, while the adjacent diagonal section, starting with the $u_2\bar{u}_1$ state, are all neutral. This pattern continues as $N_f$ gets larger. For the charged states, counting via adding the alternating diagonals, we see the pattern 
\begin{equation}
\begin{split} 
    &(N_f-1)+(N_f-3)+(N_f-5)+\cdots \\
    &=\sum_{k=1}^{n_\pm}(N_f-2k+1) =n_\pm N_f-n_\pm^2. 
\end{split}
\end{equation}
For the complex neutral states, we count
\begin{equation}
\begin{split} 
    &(N_f-2)+(N_f-4)+(N_f-6)+\cdots \\
    &=\sum_{k=1}^{n_0}(N_f-2k)=n_0N_f-n_0^2-n_0
\end{split}
\end{equation}
When $N_f$ is even, these sums terminate at $n_\pm=N_f/2$ and $n_0=N_f/2-1$ respectively, and we find that
\begin{equation}\label{evenNstates}
  \text{even:} \ \ \  N_\pm=\frac{N_f^2}{4}, \ \ \ \ N_0=\frac{N_f(N_f-2)}{4}.
\end{equation}
The total number of states include the complex conjugates of the above along with the $N_f-1$ real scalars, so that altogether, we have 
\begin{equation}\label{sumofstates}
    2N_\pm+2N_0+(N_f-1)=N_f^2-1,
\end{equation}
as required. In the case where $N_f$ is odd, evaluate (\ref{evenNstates}) for $N_f'=N_f-1$, and the remaining row/column will have $N_f-1$ total states, where half are neutral and half are charged. This pattern holds true whether the fractional charge of the remaining quark has magnitude $|Q_u|$ or $|Q_d|$. In total, we find
\begin{equation}\label{oddNstates}
    \text{odd:} \ \ \ N_{\pm}=\frac{N_f^2-1}{4}, \ \ \ \ N_0=\frac{(N_f-1)^2}{4}.
\end{equation}
It is trivial to check that (\ref{oddNstates}) satisfies (\ref{sumofstates}). 

At this stage, the number of flavors $N_f$ is arbitrary. However, we require that it satisfies an inequality related to the number of colors $N_c$ in the following way. The QCD $\beta$-function \cite{Politzer:1973fx,Gross:1973ju} is famously 
\begin{equation}
    \beta=-\frac{g^3}{16\pi^2}\Big(\frac{11}{3}N_c-\frac{2}{3}N_f\Big).
\end{equation}
Confinement requires that $\beta<0$, constraining the relationship between the number of colors and flavors to be
\begin{equation}
    \frac{11}{2}N_c>N_f.
\end{equation}
Therefore the maximum number of flavors we can consider for, say, $N_c=3$ is $N_f=16.$

\subsection{Axion Mass Spectrum}\label{sec:MassSpectrum}

The masses of the individual $\pi$-axions are related to the quark masses $m_{q_i}$, the decay constant $F_\pi$, and the dark QCD scale $\Lambda_{\text{dQCD}}$. The Gell-Mann-Oakes-Renner (GMOR) relation \cite{Gell-Mann:1968hlm} approximates the masses of the electrically neutral states as
\begin{equation}\label{GMOR}
    m_{\pi_i^0}^2\simeq\frac{\langle q\bar{q}\rangle}{F_\pi^2}\sum_i m_{q_i},
\end{equation}
where $\langle q\bar{q}\rangle\sim\Lambda_{\text{dQCD}}^3$ is the quark condensate, and the $m_{q_i}$ are the masses of the constituent quarks. We will see that it is straightforward to identify the quark content of a given $\pi$-axion state using the $SU(N_f)$ generators. 

For charged $\pi$-axions, photon loops (either from the visible or dark sector) give corrections to the GMOR relation \cite{Kaplan:2005es},
\begin{equation}
\label{eq:mpm}
m_{\pi_i^{\pm}}^2 \simeq   m_{\pi_i ^0}^2 + 2\xi_i \varepsilon^2 e^2  F_{\pi}^2,
\end{equation}
where $\xi={\cal O}(1)$. Since $F_{\pi} > 10^{11} $ GeV by assumption, if $\varepsilon={\cal O}(1)$ then the charged pions are {\it superheavy} particles, which exhibit their own interesting phenomenology \cite{Carney:2022gse}. 

\subsection{Axion-Photon Coupling}\label{sec:AxionPhotonCoupling}

We now focus on the canonical axion-photon coupling, which is the focus of many theoretical efforts \cite{Wilczek:1987mv,Kim:2008hd,Ringwald:2012hr,DiLuzio:2020wdo} and experimental searches \cite{Antypas:2022asj,Zyla:2020zbs,Adams:2022pbo,HAYSTAC:2018rwy,ADMX:2019uok,ADMX:2020ote,ADMX:2021mio,ADMX:2021nhd,Lawson:2019brd,ALPHA:2022rxj,Shaposhnikov:2023pdj} for ALPs. In our case we have,
\begin{equation}\label{interaction1}
\mathcal{L}_{\pi^I \gamma \gamma} =\frac{\lambda_I\varepsilon^2\alpha_{em}}{2 F_{\pi}}\pi_I F_{\mu\nu}\tilde{F}^{\mu\nu},
\end{equation}
where $I=1,2,..., N_f-1$ denotes the different neutral pseudoscalars. Our coupling is related to the conventional axion-photon couplings as 
\begin{eqnarray}\label{gcoupling1}
    g_{\pi \gamma \gamma} ^{(I)}= \frac{2\lambda_I\varepsilon^2\alpha_{em}}{F_\pi}.
\end{eqnarray}
The coefficients $\lambda_I$ follow from the chiral anomaly,
\begin{equation}
  \partial_\mu  j^{\mu 5 I} = -\frac{e^2}{16\pi^2}F\tilde{F}\times{\rm Tr} (T^I Q^2)
\end{equation}
where $T^I$ is an $SU(N_f)$ generator and $Q$ is the quark charge matrix  $Q= \text{diag}(Q_u,Q_d,Q_u,...) $. The chiral anomaly gets a non-zero contribution from each diagonal generator, each of which corresponds to the coupling of a particular neutral pion state, i.e.,
\begin{equation}
    \lambda_I =\frac{N_c}{4\pi}{\rm Tr}(T^I Q^2)\equiv\frac{N_c}{8\pi}\tilde{\lambda}_I.
\end{equation}
The matrices in the trace do not depend on the number of colors, allowing us to pull out a factor of $N_c$. The relevant generators for this calculation are given in (\ref{diagGens}), where $I=1,2,...,N_f-1.$ Keeping $Q_u,Q_d$ arbitrary, we find 
\begin{equation}
    \begin{split}
        \tl_1&=Q_u^2-Q_d^2, \\
        \tl_2&=-\frac{1}{\sqrt{3}}(Q_u^2-Q_d^2), \\
        \tl_3&=\sqrt{\frac{2}{3}}(Q_u^2-Q_d^2), \\
        \tl_4&=-\sqrt{\frac{2}{5}}(Q_u^2-Q_d^2), \\
        \tl_5&=\sqrt{\frac{3}{5}}(Q_u^2-Q_d^2), \\
        \tl_6&=-\sqrt{\frac{3}{7}}(Q_u^2-Q_d^2), \\
        \tl_7&=\sqrt{\frac{4}{7}}(Q_u^2-Q_d^2), \\
        \vdots
    \end{split}
\end{equation}
In order to find a closed form expression for the $\tilde{\lambda}$, note that the coefficient in front of the $(Q_u^2-Q_d^2)$ term has a numerator which is the square root of $a_n=\{1,1,2,2,3,3,...\}$, which can be represented by the sequence
\begin{equation}
    a_n=\frac{2n+1+(-1)^{n+1}}{4},
\end{equation}
while the denominator is the square root of $b_n=\{1,3,3,5,5,...\}$, which we can write as 
\begin{equation}
    b_n=\frac{2n+1+(-1)^{n}}{2},
\end{equation}
up to the alternating sign. The coefficient in front of the $n^{\text{th}}$ term is therefore $\sim (-1)^{n+1}\sqrt{a_n/b_n}$. Specifically, 
\begin{equation}\label{lambdans}
    \tl_n=\frac{(-1)^{n+1}}{\sqrt{2}}\sqrt{\frac{2n+1+(-1)^{n+1}}{2n+1+(-1)^n}}\times (Q_u^2-Q_d^2)
\end{equation}
The coefficient in front of the $N_f-1$ term is thus
\begin{equation}
    \tl_{N_f-1}=\frac{(-1)^{N_f}}{\sqrt{2}}\sqrt{\frac{2N_f-1+(-1)^{N_f}}{2N_f-1+(-1)^{N_f-1}}}\times (Q_u^2-Q_d^2).
\end{equation}
From this we find that the all $(N_f-1)$ neutral pions have roughly the same axion-photon coupling, their values differing only by ${\cal O}(1)$ factors.

\subsection{Other Portals to Standard Model}\label{sec:OtherPortals}

Finally, we consider the other possible portals  to the Standard Model. The coupling to photons can be of the form,
\begin{eqnarray}\label{interaction1}
\mathcal{L}_{\text{int}}^{(1)}&&=\frac{\lambda_1\varepsilon^2}{2 F_{\pi}}(\pi^0)F_{\mu\nu}\tilde{F}^{\mu\nu},\\
\label{interaction2}
\mathcal{L}_{\text{int}}^{(2)}&&=\frac{\lambda_2\varepsilon^2}{2}  (\pi^{+})(\pi^{-}) A_\mu A^\mu, \\
\label{interaction3}
\mathcal{L}_{\text{int}}^{(3)}&&=\frac{\lambda_3\varepsilon^2}{2 \Lambda_3 ^2}  (\pi^{+})(\pi^{-}) F_{\mu\nu}F^{\mu\nu}, \\
\label{interaction4}
\mathcal{L}_{\text{int}}^{(4)}&&=\frac{\lambda_4\varepsilon^2}{2 \Lambda_4 ^2} (\pi_i)(\pi_j) F_{\mu\nu}F^{\mu\nu}. 
\end{eqnarray}
Formally, each term is a sum over all $\pi$-axion states which participate in the interaction, which we have omitted for brevity. The first interaction (\ref{interaction1}) is the standard axion-photon coupling arising from a triangle diagram, and only the $N_f-1$ real, neutral states experience the vertex. The second interaction (\ref{interaction2}) stems from the gauge covariant derivative in scalar QED, and all charged $\pi$-axions participate in the interaction. The remaining two vertices, (\ref{interaction3}) and (\ref{interaction4}), are EFT operators arising from integrating out the heavy degrees of freedom in the dark SM. (\ref{interaction3}) entails a sum over all complex $\pi$-axions, while (\ref{interaction4}) sums over all neutral $\pi$-axions, both complex and real.

Similarly, pseudoscalar couplings to SM fermions can arise through \cite{Marsh:2015xka} 
\begin{equation}
    \mathcal{L}_{\pi N}=\frac{g_{\pi N}\varepsilon^2}{2m_N}\partial_\mu\pi_{0}(\bar{N}\gamma^\mu\gamma^5 N),
\end{equation}
\begin{equation}
\mathcal{L}_{\pi e}=
    \frac{g_{\pi e}\varepsilon^2}{2m_e}\partial_\mu\pi_{0}(\bar{e}\gamma^\mu\gamma^5 e),
\end{equation}
\begin{eqnarray}
    \mathcal{L}_{\pi N\gamma}=-\frac{i\varepsilon^2}{2M_*^2}\pi_0\bar{N}J^{\mu\nu}\gamma^5 NF_{\mu\nu},
\end{eqnarray}
where $N$ is a SM nucleon, $e$ is an electron, muon, or tau particle, $F_{\mu\nu}$ is the photon field strength, and $J^{\mu\nu}=\tfrac{i}{2}[\gamma^\mu,\gamma^\nu]$ are the Lorentz generators. Each of the couplings can be related to the decay constant $F_\pi$, namely $g_{\pi N}/(2m_N)\sim g_{\pi e}/(2m_e)\propto 1/F_\pi$, and $M_{*}\propto F_\pi$.

\section{Example: $N_f=10$}
\label{sec:Nf10}

In this section, we will turn our attention to the specific case of $N_f=10$ and $N_c=3.$ Moreover, we will additionally set the millicharge parameter to unity, $\varepsilon=1$, leaving as the only free parameters the quark masses and confinement scale. In this case the charged $\pi$-axions are superheavy, and will not be produced by misalignment \cite{Alexander:2023wgk,Maleknejad:2022gyf}.

\subsection{Spectrum of the $N_f=10$ Axiverse}\label{sec:Nf10Spectrum}

\begin{table}[h!]
Spectrum of Real Neutral $\pi$-axions in Dark QCD\\
\begin{tabular}{c|c|c}

\hline
$\pi$-axion    & quark content       & mass($m_{\pi^i} ^2$)      \\
\hline
$\pi_1$ & $u_1\bar{u}_1-d_1\bar{d}_1$ & $2m_{q} F_{\pi}$\\
  $\pi_2$      &  $u_1\bar{u}_1+d_1\bar{d}_1-2u_2\bar{u}_2$    & $3m_{q} F_{\pi}$\\
  $\pi_{3}$ & $u_1\bar{u}_1+d_1\bar{d}_1+u_2\bar{u}_2-3d_2\bar{d}_2$ & $4m_{q} F_{\pi}$\\
  $\pi_{4}$ & $\sum_{i=1}^2(u_i\bar{u}_i+d_i\bar{d}_i)-4u_3\bar{u}_3$ & $5m_{q} F_{\pi}$\\
$\pi_{5}$  & $\sum_{i=1}^2(u_i\bar{u}_i+d_i\bar{d}_i)+u_3\bar{u}_3-5d_3\bar{d}_3$  &$6m_{q} F_{\pi}$\\
$\pi_6$ & $\sum_{i=1}^3(u_i\bar{u}_i+d_i\bar{d}_i)-6u_4\bar{u}_4$ & $7m_{q} F_{\pi}$ \\
$\pi_7$ & $\sum_{i=1}^3(u_i\bar{u}_i+d_i\bar{d}_i)+u_4\bar{u}_4-7d_4\bar{d}_4$ &$8m_{q} F_{\pi}$ \\
$\pi_8$ & $\sum_{i=1}^4(u_i\bar{u}_i+d_i\bar{d}_i)-8u_5\bar{u}_5$ & $9m_{q} F_{\pi}$ \\
$\pi_9$ & $\sum_{i=1}^4 (u_i\bar{u}_i+d_i\bar{d}_i)+u_5\bar{u}_5-9d_5\bar{d}_5$ &$10m_{q} F_{\pi}$ \\ \hline 
\end{tabular}
\caption{\label{neutralmasstable}{\bf Spectrum of real, neutral $\pi$-axions}: The nine states are constructed using contractions with the diagonal generators (\ref{diagGens}). For simplicity, we are omitting the 25 charged states and 20 complex, neutral states. }
\end{table}

We will again denote each generation of dark quarks as $\{u_i,d_i\}$, $i=1,...,N_f/2$, in analogy with the SM up and down quarks.  For the example of $N_f=10,$ we have five dark quark generations, 99 total $\pi$-axion states, 9 real, neutrals, and 45 complex states along with their conjugates. Using (\ref{evenNstates}), we find that 25 of the states are charged and 20 are neutral. We have computed the 99 generators using the algorithm presented in \cite{Pfeifer2003}, which we have summarized in (\ref{diagGens}), (\ref{realoffGens}), and (\ref{imoffGens}). As is standard in a $\Sigma$ model, we can package these states in a matrix $\Sigma=\pi_IT^I,$ where $\pi_I=(\pi_1,\pi_2,...,\pi_{99})$ is the pion vector and $T^I$ are the $SU(10)$ generators. 
The result is displayed in (\ref{SigmaMat}), and the diagonal entries are given by (\ref{DiagEntries}).  

In order to make contact with the $\pi$-axion mass content, let us define some generic notation for the dark quark masses. We label a characteristic mass for each of the $N_f/2=5$ generations as $m_i,$ $i=1,...,5$, and assign a constant to the respective up and down type dark quarks in that generation, $c_{u_i}$ and $c_{d_i}$. As an example, the first generation $\{u_1,d_1\}$ have masses $\{c_{u_1}m_1,c_{d_1}m_1\}$. The SM up and down quark have masses $m_u=2.2$ MeV and $m_d=4.7$ MeV, which in this notation can be written $m_1=2.2$ MeV with $c_{u_1}=1,$ $c_{d_1}=2.14$ such that $m_u=c_{u_1}m_1$ and $m_d=c_{d_1}m_1$.

Using the GMOR relation (\ref{GMOR}) along with the corrections for the charged states (\ref{eq:mpm}), we can approximate the mass for a given $\pi$-axion in terms of the constants $c_u$, $c_d$, and the characteristic masses $m_i$. We note that all of the complex scalars are composed of one dark quark and one dark anti-quark of a different flavor, while the real scalars are composed of multiple dark quark/anti-quark pairs of the same flavor. For the states that are composed of more than three dark quark/anti-quark pairs, the GMOR relation (\ref{GMOR}) is expected to be less accurate.

 As a simple example, consider the case of completely degenerate quark masses, $m_1=m_2=...=m_5=m_q$, and $c_{u_1}=c_{d_1}=...=c_{d_5}=1$. The spectrum of pion masses is shown on the rightmost column in Table \ref{neutralmasstable}. From this one may appreciate that all 9 pions have mass within a factor of $3$: denoting $2 m_q F_{\pi} \equiv m_\pi ^2$, the masses range from $m_{\pi}$ to $\sqrt{5} m_{\pi}$. The lightest pion mass $m_{\pi}\simeq \sqrt{2 m_q F_{\pi}}$ can be expressed in terms of benchmark parameter values as
\begin{equation}
   \left( \frac{m_{\pi}}{\rm eV} \right)^2 = 2 \frac{m_q}{10^{-20} {\rm eV}} \frac{F_{\pi}}{10^{11} {\rm GeV}}.
\end{equation}
The remaining pion masses are given in Tab.~\ref{neutralmasstable}.

\begin{figure*}
    \centering
    \includegraphics[width=0.9\textwidth]{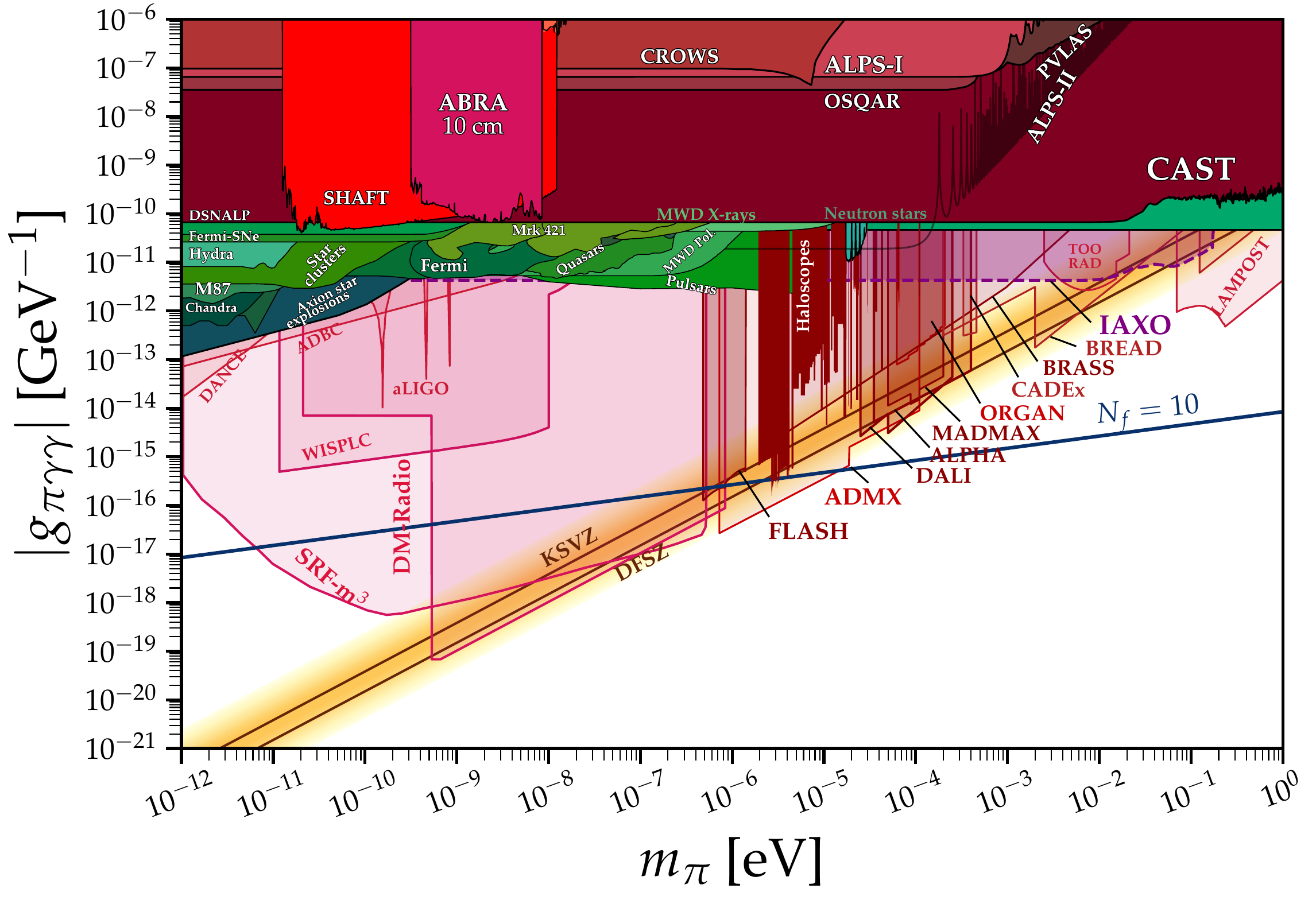}
    \caption{Constraints on the axion-photon coupling in the field theory axiverse for $SU(3)$ dark QCD with $N_f=10$ dark quarks and $\varepsilon=1$. The dark blue line indicates the model prediction for the pion-photon coupling. Here we assume the neutral pseudoscalar pions constitute the observed relic density of dark matter, and denote as $m_\pi$ the mass of the lightest pion.  }
    \label{fig:constraints}
\end{figure*}

\subsection{Experimental Constraints on axion-photon coupling}
\label{sec:Nf10Constraints}

To estimate constraints on the model, we approximate our multiple distinct resonances as a single signal. This approach differs from that taken in the context of `ALP anarchy' \cite{Chadha-Day:2023wub}, where one performs a rotation in field space to combine multiple axion-photon couplings into a single effective coupling. Our approach is justified on the basis of the tightly-packed mass spectrum, namely that all 9 fields have mass in the range $[\sqrt{2},\sqrt{10}] \sqrt{m_{q}F_{\pi}}$. 

The combined axion photon coupling then is the sum over the individual couplings, i.e. 
\begin{equation}
    \lambda_{\rm eff}\equiv \sqrt{ \sum_{I=1} ^{N_{F}-1} \lambda _I ^2}= \frac{N_c}{8\pi}\sqrt{\sum_{I=1} ^{N_F - 1}  \tl_I ^2},
\end{equation}
where the form of the $\tl_I$ are given in (\ref{lambdans}). The sum works out to be 
\begin{equation}
    \sum_{I=1}^9\tl_I^2=5\times (Q_u^2-Q_d^2)^2,
\end{equation}
in this case, thus, 
\begin{equation}
    \lambda_{\text{eff}}=\frac{\sqrt{5}N_c}{8\pi}\times (Q_u^2-Q_d^2).
\end{equation}
If we set the fractional quark charges to the analogous SM values $Q_u=2/3, Q_d=-1/3,$ along with $N_c=3$, we find a corresponding axion-photon coupling (\ref{gcoupling1}) 
\begin{equation}
    g_{\pi \gamma \gamma} = \frac{\sqrt{5}\alpha_{em}}{4\pi F_\pi}.
\end{equation}

To isolate for the mass dependence, we assume that the $\pi$-axions constitute the total dark matter relic density. That is, the relic density produced via misalignment is given by
\begin{equation}\label{misa}
    \Omega_{\pi} = \frac{1}{6} (9 \Omega_r)^{3/4} \frac{F_\pi^2}{M_{\text{Pl}}^2} \displaystyle \sum _{i} \left(\frac{m_{\pi_i}}{H_0} \right)^{1/2}  \theta_{\pi_i} ^2,
\end{equation}
where the sum is over all light stable $\pi$-axion fields, and we assume $m_{\pi^i}>10^{-28}$ eV for simplicity (see \cite{Marsh:2015xka} for the relic density when $m_{\pi^i}<10^{-28}$ eV). The left hand side of (\ref{misa}) is fixed by observation, $\Omega_{\rm DM} h^2 = 0.12$ \cite{Planck:2018vyg}.

From this we find the axion photon as a function of the lightest pion mass as,
\begin{equation}
    g_{\pi\gamma\gamma}=1.3\times 10^{-11}\alpha_{em}\theta_i\lambda_{\text{eff}}\Big(\frac{m_\pi}{\text{eV}}\Big)^{1/4} \ \text{GeV}^{-1} .
\end{equation}
where for concreteness we assume a common initial misalignment $\theta_i$ for the pions.  The constraints on this scenario are shown in Fig.~\ref{fig:constraints}.

\section{The Dark Bary-verse}\label{sec:Baryons}
Consider a color neutral state composed of $N_c$ constituent dark quarks, analogous to SM baryons,
\begin{equation}
    B\sim\prod_i^{n_u}\prod_j^{n_d}\langle u_id_j\rangle.
\end{equation}
These are color singlet states analogous to the proton and neutron, where the number of up-type $n_u$ and down-type $n_d$ valence quarks satisfy
\begin{eqnarray}\label{baryonNc}
    n_u+n_d=N_c.
\end{eqnarray}
Here we look to count the number of dark baryons as a function of the number of flavors $N_f$ and colors $N_c$. The enumeration can be understood in the following way. We can organize the counting by noting that a given state could be composed of a single type of quark, 2-types, 3-types, etc., up to all quarks being unique. Symbolically, we can write this as
\begin{eqnarray}\label{states1}
    \langle\overbrace{\underbrace{q_iq_iq_i\cdots q_iq_i}_{N_c}}^{\text{1 type}}\rangle, \ \ 
   &&  \langle \overbrace{q_i\underbrace{q_jq_j\cdots q_jq_j}_{N_c-1}}^{\text{2 types}}\rangle, \ \  \nonumber \\
    \langle \overbrace{q_iq_j\underbrace{q_kq_k\cdots q_k}_{N_c-2}}^{\text{3 types}}\rangle, \ \  
   &&  \langle\overbrace{ q_iq_jq_k\underbrace{q_l\cdots q_l}_{N_c-3}}^{\text{4 types}}\rangle, \ \ \nonumber \\
    && \vdots \nonumber \\
    && \nonumber \\
    \ \ \langle \overbrace{\underbrace{q_iq_j\cdots q_k}_{N_c-2}q_lq_l}^{N_c-1 \text{ types}}\rangle, \ \ && \langle\overbrace{\underbrace{q_iq_jq_l\cdots q_mq_n}_{N_c}}^{\text{all unique}}\rangle .
\end{eqnarray}
For states composed of a single quark type, there are obviously $N_f$ of them. For states composed of two quark types like the second bra-ket in (\ref{states1}), we have $N_f-1$ choices for the $q_i$ for each $q_j$. This totals $N_f\times\binom{N_f-1}{1}$ where $\binom{n}{k}$ is the binomial coefficient. For states composed of three quark types like the third bra-ket in (\ref{states1}), we choose the 2 $q_i,q_j$ from $N_f-1$ choices for each $q_k,$ giving $N_f\times\binom{N_f-1}{2}.$ This pattern continues until we get to the second to last bra-ket state in (\ref{states1}), where we choose $N_c-2$ states from $N_f-1$ options, for each $q_l$. This gives a contribution of $N_f\times \binom{N_f-1}{N_c-2}.$ The rightmost state is composed of all unique quarks, no repeated flavors. This contributes $\binom{N_f}{N_c}$. Adding all of these possibilities together gives the total number of baryon states built from a unique composition of quarks:
\begin{equation}
    N_B(N_f,N_c)=\binom{N_f}{N_c}+N_f\times\sum_{k=0}^{N_c-2}\binom{N_f-1}{k}.
\end{equation}
The example of $N_f=10$ and $N_c=3$ results in the total number of baryons being $N_B(10,3)=220.$

Notably, this result does not count states that have the same quark composition but different total angular momentum $J$. As an example from the Standard Model, the proton $p^+$ and $\Delta^+$ particle are both composed of $uud$ valence quarks, but the proton has $J=1/2$ while the $\Delta^+$ has $J=3/2.$ They of course have the same charge, but they have different masses and radically different lifetimes. We further note that the charge of a given baryon $B$ is straightforwardly obtained by adding the fractional charge of its quark content, $Q_B=Q_un_u+Q_dn_d,$ where the up-type and down-type quark charges satisfy $Q_u-Q_d=\varepsilon.$

\section{Discussion}
\label{sec:Discussion}

In this article we have presented a \textit{field theory axiverse} by assuming a simple, dark QCD with ultralight quarks and a large dark confinement scale, satisfying the hierarchy $m_q/\Lambda_{\text{dQCD}}\ll 1.$ The relic dark matter candidates are the PNGBs associated to the spontaneous breaking $SU(N_f)_{\text{L}}\times SU(N_f)_{\text{R}}\rightarrow SU(N_f)_{\text{V}}$, of which there are $N_f^2-1$ states. The dark quarks are (milli)charged under $U(1)_{\text{EM}}$, resulting in a portal between the dark sector and the SM photon. The $N_f^2-1$ PNGBs are either charged or neutral, complex scalar fields, or real, neutral scalars analogous to the SM charged pion, kaon, and neutral pion, respectively.

The spectrum and couplings of the field theory axiverse follows from simple considerations, and indeed in Sec. \ref{sec:Enumerating} we derived formulae which counts the number of unique states that are real pseudoscalars or complex pseudoscalars, and whether the states are charged or neutral for the latter case. The formulae apply for arbitrary $N_f$, provided the number of flavors satisfies the inequality $N_f<\tfrac{11}{2}N_c$ such that the $\beta$-function is negative and the quarks are confined. The masses are then dictated by the the GMOR relation approximating the $\pi$-axion masses, and the couplings by the chiral anomaly.

As an concrete example of the power of this approach, we have considered the specific case of $N_f=10$ and $N_c=3$, where we find 9 real, 25 charged, and 20 complex neutral states, totaling (along with the conjugates) $100^2-1=99$ $\pi$-axions. Given that the 9 neutral pi-axions all have mass within the range $\sqrt{2}\sqrt{m_{q}F_{\pi}}$ to $\sqrt{10} \sqrt{m_{q}F_{\pi}}$, we estimated the experimental constraints on this scenario by approximating the distinct resonances as a single signal, combining the 9 individual couplings into an effective coupling.

Assuming that the relic density is produced by misalignment and that the $\pi$-axions constitute the total dark matter density today, we produced the current and projected future constraints Fig. \ref{fig:constraints}, illustrating both the contrast between the field theory axiverse and the standard KSVZ and DFSZ axions, as well as the regimes where the model is still in play.

An intriguing feature of this path to the axiverse is the prediction of a corresponding dark {\it bary-verse}, namely the stable baryonic states of the dark QCD theory. While the axiverse is naturally on the ultralight end of the mass spectrum, the dark baryons are naturally {\it superheavy}. Superheavy dark matter brings its own rich phenomenology \cite{Carney:2022gse}, adding to the opportunities for testing this scenario. Analogous to the axion-like states, in this work we have derived a formula for counting the dark baryon-like states in the field theory axiverse, where the number of up- and down-type valence quarks satisfy $n_u+n_d=N_c,$ finding a total of 220 states for $N_f=10,N_c=3$.

Our results present a number of interesting directions for future work. Top amongst the list is a dedicated analysis of broadband direct detection searches for multicomponent axion dark matter arising in this scenario, and other phenomenological aspects of multi-axion scenarios (see e.g.~\cite{Neves:2024rju,Chadha-Day:2023wub,Stott:2017hvl,Stott:2018opm,Mehta:2021pwf}), such as axion stars \cite{Visinelli:2017ooc,Amin:2020vja,Amin:2021tnq,Du:2023jxh,Escudero:2023vgv,Chung-Jukko:2023cow}. It will also be interesting to consider  fuzzy dark matter phenomenology such as vortices \cite{Hui:2020hbq}, and other substructure \cite{Alexander:2019qsh}, which can leave an imprint in e.g., strong gravitational lensing \cite{Alexander:2019puy} or  cosmic filaments \cite{Alexander:2021zhx}. Ultralight axion-like particles can also play a role in cosmological parameter tensions (see e.g.~\cite{Rogers:2023ezo}). It remains an interesting question if these observable windows can be used to discriminate between axion-like particle candidates, in particular the pions presented here vs. other composite ultralight dark matter models such as \cite{Alexander:2020wpm} and \cite{Alexander:2018fjp}. Complementary to these studies would be a dedicated analysis of the superheavy dark baryons of the dark QCD theory and their associated phenomenology.

Finally, it will also be interesting to embed this scenario into other models, such as a Grand Unified Theory with an $SO(10)$ gauge group. This is a natural context where one might expect a dark strong force with $N_f \gtrsim 10$ charged fermions. An interesting question is the pathway to discriminating the dark Standard Model construction of \cite{Alexander:2023wgk} from a dark GUT and from the minimal field theory axiverse presented here. We leave this and other directions for future work.\\

\begin{center}
{\bf ACKNOWLEDGEMENTS }
\end{center}
 The authors thank Heliudson Bernardo, Humberto Gilmer, Michael Toomey, and Wenzer Qin for helpful comments.  S.A. and T.M. are supported by the Simons Foundation, Award 896696. E.M. is supported in part by a Discovery Grant from the Natural Sciences and Engineering Research Council of Canada, and by a New Investigator Operating Grant from Research Manitoba.

\vspace{2cm}

\appendix

\section{$SU(N)$ generators}\label{app:SUNgennies}

An algorithm for constructing the $N^2-1$ generators for arbitrary $N$ was described in \cite{Pfeifer2003}. We note that the labelling in this approach differs from the standard labeling of the $SU(N)$ generators. For example the $SU(3)$ generators (Gell-Mann matrices \cite{Gell-Mann:1962yej}) are organized in increasing order of $SU(2)$ subgroups, i.e. $\{\lambda_1,\lambda_2,\lambda_3\}$ and $\{\lambda_4,\lambda_5, a\lambda_3+b\lambda_8\}$ form two of the three $SU(2)$ subalgebras of $SU(3)$, where $a$ and $b$ are constants. Although the labeling is not important, our notation implies, for example, that the analog of the SM neutral pion, usually identified as $\pi_3$ in the $\Sigma$-model, is our $\pi_1$.

The following matrices, which we will call $t^I$, satisfy $[t^I,t^J]=2\delta^{IJ}$. These are related to those in the main text by $T^I=\tfrac{1}{2}t^I$, such that we adhere to the standard physics convention for the index of the fundamental representation being 1/2, i.e. $[T^I,T^J]=\tfrac{1}{2}\delta^{IJ}$.  The $N-1$ diagonal generators are given by 
\begin{widetext}
\begin{equation}\label{diagGens}
    \begin{pmatrix}
        1&&&&& \\
        &-1 &&&& \\
        &&0&&& \\
        &&&0&& \\ 
        &&&&\ddots& \\
        &&&&& 0
    \end{pmatrix}, \ \ \frac{1}{\sqrt{3}}\begin{pmatrix}
        1&&&&& \\
        &1 &&&& \\
        &&-2&&& \\
        &&&0&& \\ 
        &&&&\ddots& \\
        &&&&& 0
    \end{pmatrix}, ..., \ \ \sqrt{\frac{2}{N(N-1)}}\begin{pmatrix}
        1&&&&& \\
        &1 &&&& \\
        &&1&&& \\
        &&&\ddots&& \\ 
        &&&&1& \\
        &&&&& -N+1
    \end{pmatrix},
\end{equation}
while the remaining $N(N-1)$ generators are constructed from the $N(N-1)/2$ real matrices 
\begin{equation}\label{realoffGens}
    \begin{pmatrix}
        0&1&&&&0 \\
        1&0 &&&& \\
        &&&&& \\
        &&&&& \\ 
        &&&&\ddots& \\
        0&&&&& 0
    \end{pmatrix}, \ \ \ \begin{pmatrix}
        0&0&1&&&0 \\
        0&0 &&&& \\
        1&&0&&& \\
        &&&&& \\ 
        &&&&\ddots& \\
        0&&&&& 0
    \end{pmatrix}, ..., \ \ \begin{pmatrix}
        0&0&&&&0 \\
        0&0 &&&& \\
        &&\ddots&&& \\
        &&&\ddots&& \\ 
        &&&&0&1 \\
        0&&&&1& 0
    \end{pmatrix},
\end{equation}
    and $N(N-1)/2$ complex matrices 

    \begin{equation}\label{imoffGens}
    \begin{pmatrix}
        0&-i&&&&0 \\
        i&0 &&&& \\
        &&&&& \\
        &&&&& \\ 
        &&&&\ddots& \\
        0&&&&& 0
    \end{pmatrix}, \ \ \ 
    \begin{pmatrix}
        0&0&-i&&&0 \\
        0&0 &&&& \\
        i&&0&&& \\
        &&&&& \\ 
        &&&&\ddots& \\
        0&&&&& 0
    \end{pmatrix}, ..., \ \ \begin{pmatrix}
        0&0&&&&0 \\
        0&0 &&&& \\
        &&\ddots&&& \\
        &&&\ddots&& \\ 
        &&&&0&-i \\
        0&&&&i& 0
    \end{pmatrix}.
\end{equation}

\section{$SU(10)$ $\pi$-axion $\Sigma$-matrix}\label{app:SigmaMatrix}
Using the algorithm to construct the $10^2-1=99$ generators for $SU(10)$, we compute the contraction with the pion vector $\pi_I$ to calculate $\Sigma=t^I\pi_I$, shown in (\ref{SigmaMat}).

\begin{equation}\label{SigmaMat}
\left(
\begin{array}{cccccccccc}
 \pi_A& \pi _{10}-i \pi _{55} & \pi _{11}-i \pi _{56} & \pi _{13}-i \pi _{58} & \pi _{16}-i \pi _{61} & \pi _{20}-i \pi _{65} & \pi _{25}-i \pi _{70} & \pi _{31}-i \pi _{76} & \pi _{38}-i \pi _{83} & \pi _{46}-i \pi _{91} \\
 \pi _{10}+i \pi _{55} & \pi_B & \pi _{12}-i \pi _{57} & \pi _{14}-i \pi _{59} & \pi _{17}-i \pi _{62} & \pi _{21}-i \pi _{66} & \pi _{26}-i \pi _{71} & \pi _{32}-i \pi _{77} & \pi _{39}-i \pi _{84} & \pi _{47}-i \pi _{92} \\
 \pi _{11}+i \pi _{56} & \pi _{12}+i \pi _{57} & \pi_C & \pi _{15}-i \pi _{60} & \pi _{18}-i \pi _{63} & \pi _{22}-i \pi _{67} & \pi _{27}-i \pi _{72} & \pi _{33}-i \pi _{78} & \pi _{40}-i \pi _{85} & \pi _{48}-i \pi _{93} \\
 \pi _{13}+i \pi _{58} & \pi _{14}+i \pi _{59} & \pi _{15}+i \pi _{60} & \pi_D & \pi _{19}-i \pi _{64} & \pi _{23}-i \pi _{68} & \pi _{28}-i \pi _{73} & \pi _{34}-i \pi _{79} & \pi _{41}-i \pi _{86} & \pi _{49}-i \pi _{94} \\
 \pi _{16}+i \pi _{61} & \pi _{17}+i \pi _{62} & \pi _{18}+i \pi _{63} & \pi _{19}+i \pi _{64} & \pi_E & \pi _{24}-i \pi _{69} & \pi _{29}-i \pi _{74} & \pi _{35}-i \pi _{80} & \pi _{42}-i \pi _{87} & \pi _{50}-i \pi _{95} \\
 \pi _{20}+i \pi _{65} & \pi _{21}+i \pi _{66} & \pi _{22}+i \pi _{67} & \pi _{23}+i \pi _{68} & \pi _{24}+i \pi _{69} &\pi_F & \pi _{30}-i \pi _{75} & \pi _{36}-i \pi _{81} & \pi _{43}-i \pi _{88} & \pi _{51}-i \pi _{96} \\
 \pi _{25}+i \pi _{70} & \pi _{26}+i \pi _{71} & \pi _{27}+i \pi _{72} & \pi _{28}+i \pi _{73} & \pi _{29}+i \pi _{74} & \pi _{30}+i \pi _{75} & \pi_G & \pi _{37}-i \pi _{82} & \pi _{44}-i \pi _{89} & \pi _{52}-i \pi _{97} \\
 \pi _{31}+i \pi _{76} & \pi _{32}+i \pi _{77} & \pi _{33}+i \pi _{78} & \pi _{34}+i \pi _{79} & \pi _{35}+i \pi _{80} & \pi _{36}+i \pi _{81} & \pi _{37}+i \pi _{82} & \pi_H& \pi _{45}-i \pi _{90} & \pi _{53}-i \pi _{98} \\
 \pi _{38}+i \pi _{83} & \pi _{39}+i \pi _{84} & \pi _{40}+i \pi _{85} & \pi _{41}+i \pi _{86} & \pi _{42}+i \pi _{87} & \pi _{43}+i \pi _{88} & \pi _{44}+i \pi _{89} & \pi _{45}+i \pi _{90} & \pi_I & \pi _{54}-i \pi _{99} \\
 \pi _{46}+i \pi _{91} & \pi _{47}+i \pi _{92} & \pi _{48}+i \pi _{93} & \pi _{49}+i \pi _{94} & \pi _{50}+i \pi _{95} & \pi _{51}+i \pi _{96} & \pi _{52}+i \pi _{97} & \pi _{53}+i \pi _{98} & \pi _{54}+i \pi _{99} & \pi_J \\
\end{array}
\right)
\end{equation}
The diagonal terms are explicitly
\begin{equation}\label{DiagEntries}
    \begin{split}
        \pi_A&=\pi _1+\frac{\pi _2}{\sqrt{3}}+\frac{\pi _3}{\sqrt{6}}+\frac{\pi _4}{\sqrt{10}}+\frac{\pi _5}{\sqrt{15}}+\frac{\pi _6}{\sqrt{21}}+\frac{\pi _7}{2 \sqrt{7}}+\frac{\pi _8}{6}+\frac{\pi _9}{3 \sqrt{5}}, \\
        \pi_B&=-\pi _1+\frac{\pi _2}{\sqrt{3}}+\frac{\pi _3}{\sqrt{6}}+\frac{\pi _4}{\sqrt{10}}+\frac{\pi _5}{\sqrt{15}}+\frac{\pi _6}{\sqrt{21}}+\frac{\pi _7}{2 \sqrt{7}}+\frac{\pi _8}{6}+\frac{\pi _9}{3 \sqrt{5}}, \\
        \pi_C&=-\frac{2 \pi _2}{\sqrt{3}}+\frac{\pi _3}{\sqrt{6}}+\frac{\pi _4}{\sqrt{10}}+\frac{\pi _5}{\sqrt{15}}+\frac{\pi _6}{\sqrt{21}}+\frac{\pi _7}{2 \sqrt{7}}+\frac{\pi _8}{6}+\frac{\pi _9}{3 \sqrt{5}}, \\
        \pi_D&=-\sqrt{\frac{3}{2}} \pi _3+\frac{\pi _4}{\sqrt{10}}+\frac{\pi _5}{\sqrt{15}}+\frac{\pi _6}{\sqrt{21}}+\frac{\pi _7}{2 \sqrt{7}}+\frac{\pi _8}{6}+\frac{\pi _9}{3 \sqrt{5}}, \\
        \pi_E&= -2 \sqrt{\frac{2}{5}} \pi _4+\frac{\pi _5}{\sqrt{15}}+\frac{\pi _6}{\sqrt{21}}+\frac{\pi _7}{2 \sqrt{7}}+\frac{\pi _8}{6}+\frac{\pi _9}{3 \sqrt{5}}, \\
        \pi_F&= -\sqrt{\frac{5}{3}} \pi _5+\frac{\pi _6}{\sqrt{21}}+\frac{\pi _7}{2 \sqrt{7}}+\frac{\pi _8}{6}+\frac{\pi _9}{3 \sqrt{5}} ,\\
        \pi_G&=-2 \sqrt{\frac{3}{7}} \pi _6+\frac{\pi _7}{2 \sqrt{7}}+\frac{\pi _8}{6}+\frac{\pi _9}{3 \sqrt{5}}, \\
        \pi_H&= -\frac{1}{2} \sqrt{7} \pi _7+\frac{\pi _8}{6}+\frac{\pi _9}{3 \sqrt{5}}, \\
        \pi_I&= \frac{\pi _9}{3 \sqrt{5}}-\frac{4 \pi _8}{3},\\
        \pi_J&=-\frac{3 \pi _9}{\sqrt{5}}.
    \end{split}
\end{equation}

\end{widetext}

 \bibliographystyle{JHEP}
\bibliography{refs.bib}

\end{document}